\documentstyle[epsfig,twoside,rotating]{aa}
\newcommand{\Msol}{M_\odot}

\newcommand{\degree}{^\circ}

\begin{document}
\title{Observation of Microlensing towards the
Galactic Spiral Arms. 
{\sc EROS II} 2 year survey
\thanks{This work is based on observations made at the
European Southern Observatory, La Silla, Chile.}
}
\author{{\sc EROS} Collaboration}

\author{
F.~Derue\inst{1}, 
C.~Afonso\inst{2},
C.~Alard\inst{3},
J-N.~Albert\inst{1},
A.~Amadon\inst{2},
J.~Andersen\inst{4},
R.~Ansari\inst{1}, 
\'E.~Aubourg\inst{2}, 
P.~Bareyre\inst{5}, 
F.~Bauer\inst{2},
J-P.~Beaulieu\inst{6},
A.~Bouquet\inst{5},
S.~Char\inst{7}\thanks{deceased},
X.~Charlot\inst{2},
F.~Couchot\inst{1}, 
C.~Coutures\inst{2}, 
R.~Ferlet\inst{6},
J-F.~Glicenstein\inst{2},
B.~Goldman\inst{2},
A.~Gould\inst{8}\thanks{Alfred P. Sloan Foundation Fellow},
D.~Graff\inst{2,9},
M.~Gros\inst{2}, 
J.~Ha\"{\i}ssinski\inst{1}, 
J-C.~Hamilton\inst{5},
D.~Hardin\inst{2},
J.~de Kat\inst{2},
A.~Kim\inst{5},
T.~Lasserre\inst{2},
\'E.~Lesquoy\inst{2},
C.~Loup\inst{6},
C.~Magneville \inst{2}, 
B.~Mansoux\inst{1}, 
J-B.~Marquette\inst{6},
\'E.~Maurice\inst{10}, 
A.~Milsztajn \inst{2},  
M.~Moniez\inst{1},
N.~Palanque-Delabrouille\inst{2},
O.~Perdereau\inst{1},
L.~Pr\'evot\inst{10},
N.~Regnault\inst{1},
J.~Rich\inst{2}, 
M.~Spiro\inst{2},
A.~Vidal-Madjar\inst{6},
L.~Vigroux\inst{2},
S.~Zylberajch\inst{2}
\\   \indent   \indent
The {\sc EROS} collaboration\\
{\it in memory of Sergio Char}
}
\institute{
Laboratoire de l'Acc\'{e}l\'{e}rateur Lin\'{e}aire,
{\sc IN2P3-CNRS}, Universit\'e de Paris-Sud, B.P. 34, 91898 Orsay Cedex, France
\and
{\sc CEA}, {\sc DSM}, {\sc DAPNIA},
Centre d'\'Etudes de Saclay, 91191 Gif-sur-Yvette Cedex, France
\and
{\sc DASGAL}, {\sc INSU-CNRS}, 77 avenue de l'Observatoire, 75014 Paris, France
\and
Astronomical Observatory, Copenhagen University, Juliane Maries Vej 30, 
2100 Copenhagen, Denmark
\and
Coll\`ege de France, {\sc LPCC}, {\sc IN2P3-CNRS}, 
11 place Marcellin Berthelot, 75231 Paris Cedex, France
\and
Institut d'Astrophysique de Paris, {\sc INSU-CNRS},
98~bis Boulevard Arago, 75014 Paris, France
\and
Universidad de la Serena, Facultad de Ciencias, Departamento de Fisica,
Casilla 554, La Serena, Chile
\and
Department of Astronomy, Ohio State University, Columbus, Ohio 43210, U.S.A.
\and
Department of Physics, Ohio State University, Columbus, Ohio 43210, U.S.A.
\and
Observatoire de Marseille, {\sc INSU-CNRS},
2 place Le Verrier, 13248 Marseille Cedex 04, France
}

\offprints{M. Moniez: Moniez@lal.in2p3.fr; \\
{\it see also our WWW server at  URL :} \\
{\tt http://www.lal.in2p3.fr/EROS/eros.html}}
\date{Received 13/03/1999, accepted }

\thesaurus{10.08.1;10.11.1;10.19.2;10.19.3;12.04.1;12.07.1}

\maketitle
  \markboth{F. Derue et al. : Observation of Microlensing towards the
Galactic Spiral Arms}{F. Derue et al. : 
Observation of Microlensing towards the
Galactic Spiral Arms}

\begin{abstract}
We present the analysis of the light curves of
8.5 million stars observed during two seasons
by {\sc EROS} (Exp\'erience de Recherche
d'Objets Sombres), in the Galactic plane away from the bulge.
Three stars have been found that exhibit luminosity variations
compatible with gravitational microlensing effects due to
unseen objects.
The corresponding optical depth, averaged over four directions,
is $\bar\tau = 0.38^{+0.53}_{-0.15} \times 10^{-6}$.
All three candidates have long Einstein radius crossing times
($\sim$ 70 to 100 days).
For one of them,
the lack of evidence for a parallax or a source size effect
enabled us to constrain the lens-source 
configuration.
Another candidate displays a modulation
of the magnification, which is compatible with the lensing
of a binary source.

The interpretation of the optical depths inferred from these
observations is hindered by the imperfect knowledge of
the distance to the target stars. Our measurements are
compatible with expectations from simple galactic models
under reasonable assumptions on the target distances.
\end{abstract}
\keywords{Galaxy: halo -- Galaxy: kinematics and dynamics -- Galaxy: stellar content -- Galaxy: structure -- {\itshape (Cosmology:)} gravitational lensing 
\section{Introduction}
Since the seminal paper of Bohdan Paczy\'nski \cite{pacz},
observations have demonstrated that
gravitational microlensing is an efficient tool to
investigate the Milky Way structure. 
After the first detections of microlensing effects
towards the Large Magellanic Cloud (\cite{machlmc}; \cite{eroslmc})
and towards the Galactic bulge (\cite{oglpr}; \cite{machobulbe}), 
searches for microlensing have entered an active era.
The results of recent campaigns of observations
are somewhat difficult to interpret.
The negative search for short duration events and the
rarity of long duration events found towards the
Magellanic Clouds (\cite{Eros1LMC}; \cite{MachoLMC-2ans};
\cite{ErosSMC} and \cite{ErosMacho}) imply
that only heavy dark compact objects ($M>10^{-2}\Msol$)
could account for a significant fraction ($>_{_{\!\!\!\!\!\sim}} 25\%$)
of the halo mass required to
explain the rotation curve of our galaxy. 
\begin{figure}
\begin{center}
\mbox{\epsfig{file=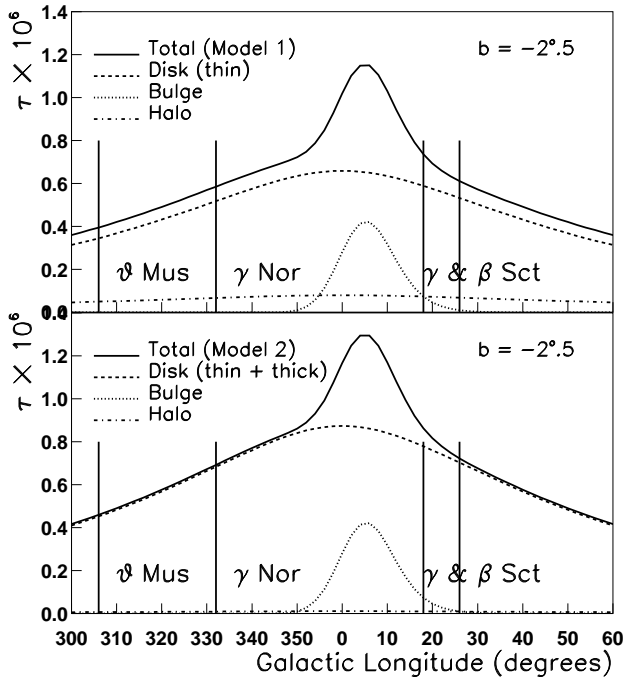,width=9.cm,angle=0.}}
\caption[]{Expected optical depth ($\times 10^{6}$) up to $7\ \rm{kpc}$
for the different components of the Milky Way as a function of
the Galactic longitude at $b= -2.\hskip-2pt \degree 5$ for two models
of the Galaxy (see Table 4).
The 4 directions towards the spiral arms are indicated.
Notice that at this Galactic
latitude the bulge and disc contributions are already reduced by
a factor $\simeq 2$ with respect to zero latitude; moreover,
the bulge contribution also changes dramatically with the distance to
the target (here assumed to be at $7\ \rm{kpc}$, while the Galactic Centre
is at $8.5\ \rm{kpc}$).}
\label{optvsl}
\end{center}
\end{figure}
On the other hand, the optical depth measured in the 
Galactic bulge direction (\cite{oglgc}; \cite{MachoBulge-1an})
appears to be significantly larger than expected from
the contributions of the disc, the bulge and the halo.
The latter measurements have led to the hypothesis of an
ellipsoidal structure of the Galactic bulge.
It is thus important to disentangle the different
contributions to this large optical depth.
The estimate of the disc contribution can be refined
by investigating lines of sight that do
not go through the hypothetic ellipsoidal structure (see Fig. \ref{optvsl}).
Therefore the {\sc EROS} team has chosen to search
for microlensing not only towards the Magellanic Clouds and
the Galactic bulge, but also
in four regions of the Galactic plane,
located at a large angle from the Galactic Centre.
\section{The observations}
\subsection{Data taking}
Since July 1996 the {\sc EROS} team has been using
at La Silla observatory the {\sc MARLY} telescope
(1 m, f/5) with a dichroic beam-splitter allowing simultaneous
imaging of a
$0.\hskip-2pt \degree 7 (\alpha) \times 1.\hskip-2pt \degree 4 (\delta)$
field in {\sc EROS}-visible and {\sc EROS}-red wide passbands.
Photons are collected by two cameras, each equipped with a $2\times 4$
mosaic of $2K\times 2K$ LORAL CCDs (\cite{TechniqueBauer}).
In this analysis, seven sub-fields are considered per image because
one of our 16 CCDs is not operational.
The pixel size is $0.6$ arcsec, and the
median global seeing is $2$ arcsec.
CCDs are read-out in parallel in approximately 50 s, during
which the telescope 
moves towards the next field to be imaged.
After acquisition by the VME system, raw data are reduced
by DEC-Alpha workstations using flat-field images
taken at the beginning of the night. The
DLT tapes produced are shipped to the {\sc CCPN}
({\sc IN2P3} computing centre, {\sc CNRS}) in Lyons, France, for subsequent
processing.

The two {\sc EROS} passbands are nonstandard.
{\sc EROS}-red passband is centered on $\bar\lambda = 762\ {\rm nm}$
with a full width half maximum $\Delta\lambda \simeq 200\ {\rm nm}$,
and {\sc EROS}-visible passband is centered on $\bar\lambda = 600\ {\rm nm}$
with  $\Delta\lambda \simeq 200\ {\rm nm}$.
Our calibration studies show that the corresponding magnitudes
$R_{EROS}$ and $V_{EROS}$ are close to Cousins I and Johnson R 
to $\pm 0.3$ magnitudes. 
These $R_{EROS}$ and $V_{EROS}$ magnitudes will be used all 
along this article.
\subsection{The targets}
\label{targets}
\begin{table}[h!] 
\begin{center}
\caption{Description of 
the 29 fields monitored in the spiral arms program. 
This table gives the fields centres, the 
averaged number of measurements and the number of light curves
analysed in this article. Field gn401 has not been studied yet.}
\label{tab1}
\begin{tabular}{|c|c|c|c|c|}\hline
\multicolumn{1}{|c}{Field }&
\multicolumn{1}{|c|}{$\alpha$ (h:m:s)}&
\multicolumn{1}{|c|}{$\delta$ (d:m:s)}&\# of & stars\\ 
number& J2000 & J2000 & meas.& (million)\\ \hline
\multicolumn{3}{|c|} { Scutum ($\beta$ Sct)} &
53 & 1.96 \\ \hline		   
bs300 &  18:43:22.0 & -07:40:53  & 55 & 0.33\\        
bs301 &  18:43:27.0 & -06:13:42  & 52 & 0.23\\        
bs302 &  18:46:16.0 & -07:22:45  & 50 & 0.32 \\         
bs303 &  18:46:20.0 & -05:55:35  & 50 & 0.28 \\         
bs304 &  18:49:21.0 & -06:45:51  & 54 & 0.43 \\        
bs305 &  18:52:26.0 & -06:35:44  & 53 & 0.37 \\ \hline  

\multicolumn{3}{|c|}{Scutum  ($\gamma$ Sct)} & 51 & 1.70 \\ \hline
gs200 &  18:28:03.0 & -14:51:06 & 55 & 0.32\\        
gs201 &  18:31:15.0 & -14:14:38 & 52 & 0.30\\        
gs202 &  18:31:33.0 & -12:48:53 & 49 & 0.29\\        
gs203 &  18:34:22.0 & -14:31:39 & 50 & 0.34\\         
gs204 &  18:34:28.0 & -13:04:31 & 50 & 0.45\\ \hline 
									   
\multicolumn{3}{|c|} {Norma ($\gamma$ Nor)}  & 100 & 3.01 \\ \hline 
		   
gn400 &  16:09:45.0 & -53:07:03  & 100 & 0.37 \\         
gn401 &  16:18:22.0 & -51:44:43  & -  & - \\         
gn402 &  16:14:57.0 & -53:04:35  & 101 &0.25 \\
gn403 &  16:22:28.0 & -52:06:20  & 108 &0.30\\         
gn404 &  16:19:09.0 & -53:26:38  & 106 &0.29\\         
gn405 &  16:26:52.0 & -52:21:02  & 111 &0.22\\         
gn406 &  16:23:54.0 & -53:43:53  & 106 &0.33\\         
gn407 &  16:31:31.0 & -52:28:44  & 107 &0.22\\         
gn408 &  16:28:42.0 & -53:51:58  & 92 &0.22\\         
gn409 &  16:15:51.0 & -54:48:45  & 90 &0.26\\         
gn410 &  16:20:30.0 & -55:04:18  & 82 &0.23\\         
gn411 &  16:09:37.0 & -55:10:07  & 90 &0.32\\ \hline  
									   
\multicolumn{3}{|c|} { Musca ($\theta$ Mus)} & 66 & 1.77 \\ \hline  
               
tm500 &  13:27:04.0 & -63:02:18 & 65 &0.31 \\          
tm501 &  13:31:18.0 & -63:34:41 & 64 &0.33 \\ 	   
tm502 &  13:34:52.0 & -64:10:30 & 68 &0.33 \\  	   
tm503 &  13:23:58.0 & -64:59:52 & 64 &0.29 \\          
tm504 &  13:12:12.0 & -64:06:49 & 70 &0.23\\          
tm505 &  13:16:15.0 & -64:40:50 & 65 & 0.28 \\ \hline
\multicolumn{4}{|r|} { Total} & 8.44 \\ \hline  
  
\end{tabular}
\end{center}
\end{table}
Four different directions are monitored in the Galactic plane, 
away from the bulge, totalizing 29 fields
which have a high stellar density, and
cover a wide range of Galactic longitude.
We refer to them as $\beta$ \& $\gamma$ Sct, $\gamma$ Nor and $\theta$ Mus. 
Table \ref{tab1} gives the coordinates 
and available data related to the monitored directions;
Fig. \ref{fg1} displays the positions of the fields in 
Galactic coordinates and 
Fig. \ref{sampling} shows the observation periods and average time 
sampling.
The exposure times of 2 and 3 minutes were chosen to
optimize the global sensitivity of the photometric measurements
taken during microlensing magnifications
(a compromise between the number of measurements and their
precision, see \cite{Mansoux}).
\begin{figure*}
\begin{center}
\mbox{\epsfig{file=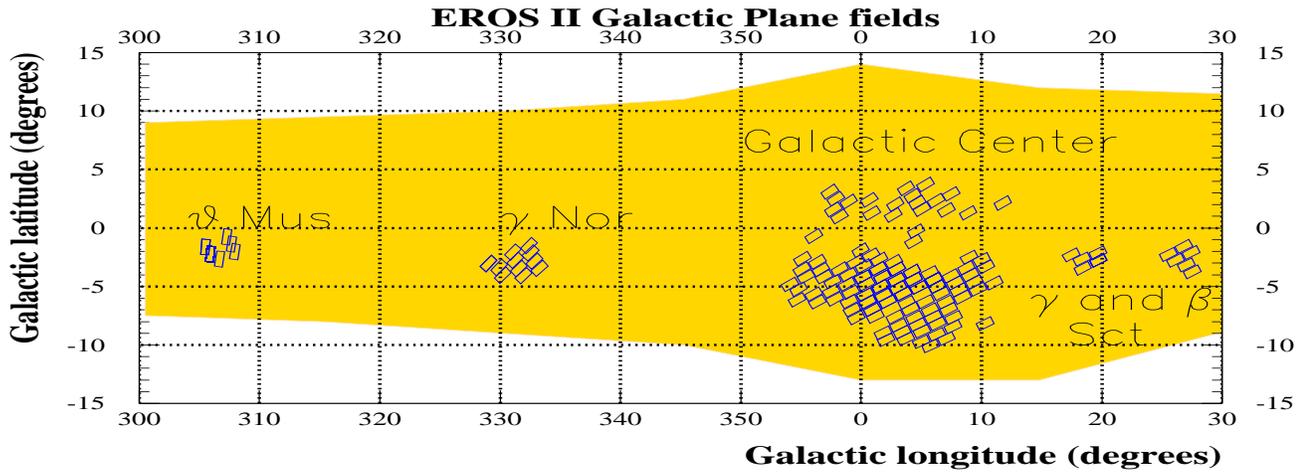,width=18cm,height=7cm}}
\caption[]{Map of the Galactic plane fields (Galactic coordinates)
monitored by {\sc EROS} for the microlensing search.
The shaded area represents the shape of the Galaxy. 
We have indicated our Galactic Centre fields and the 4 
directions towards the arms.}
\label{fg1}
\end{center}
\end{figure*}
\begin{figure}
\begin{center}
\mbox{\epsfig{file=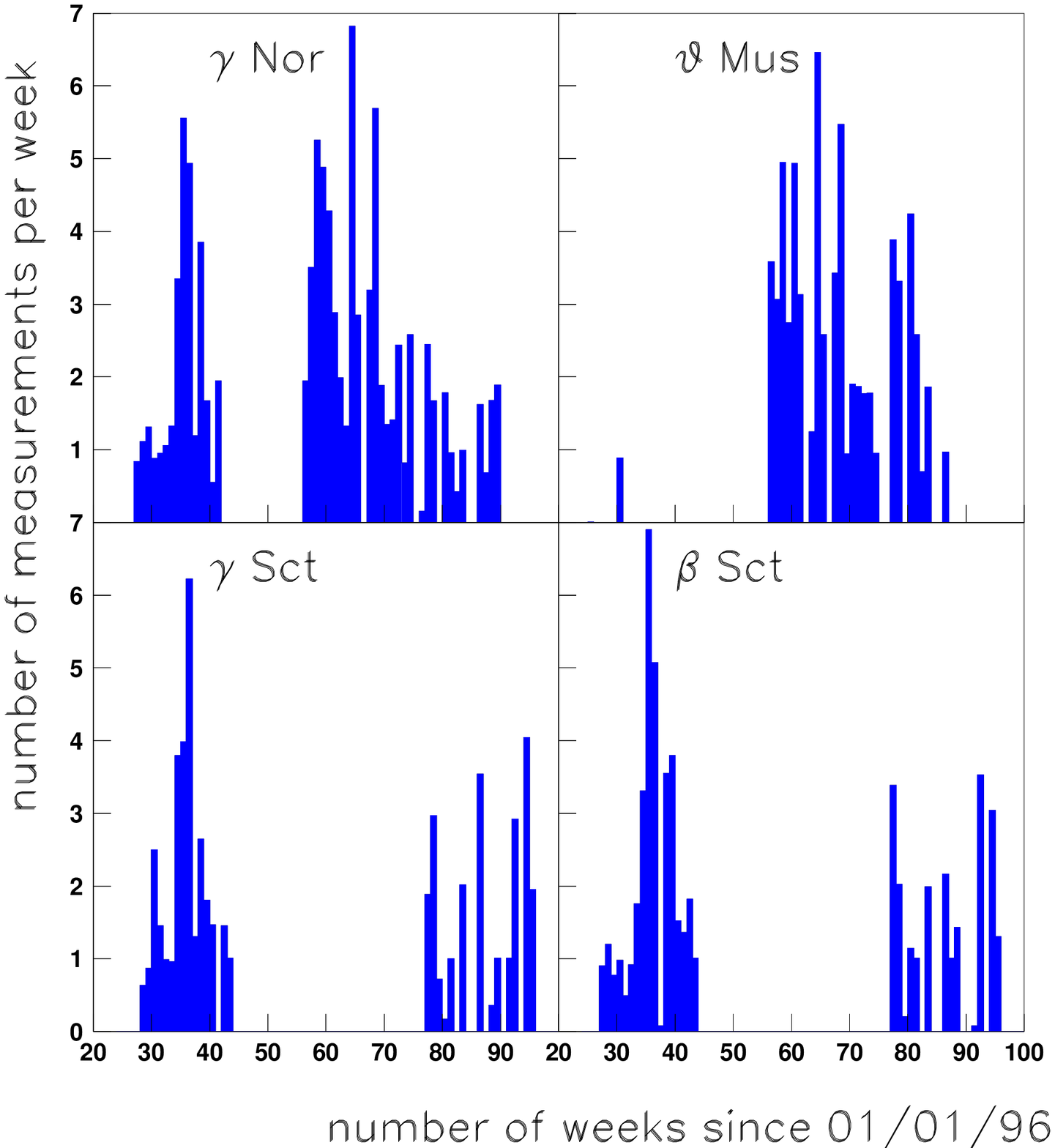,width=9.5cm}}
\caption[]{Time sampling for each direction monitored towards the 
spiral arms, in number of measurements per week.
$\gamma$ Nor is monitored between January and October, 
$\theta$ Mus is observed between January and August. 
$\beta$ \& $\gamma$ Sct are monitored between May and November.}
\label{sampling}
\end{center}
\end{figure}

By contrast with the Magellanic Clouds,
the distance distribution of the monitored stars is imperfectly known,
and should {\it a priori} vary with the limiting magnitude.
In our detection conditions, the populations of stars used
to obtain the optical depths given in Sect. 4.1 below are those
described by the colour-magnitude diagrams of Fig. \ref{diagHR}.
An analysis of these diagrams has shown that their content is
dominated by a population of source stars located $\sim 7$ kpc away,
undergoing an interstellar extinction 
of about 3 magnitudes (see \cite{Mansoux} for more details).
This distance estimate is in rough agreement with the distance to the
spiral arms deduced from \cite{Georgelin} and \cite{Russeil}, 
and will be used in this paper.
\begin{figure*}
\begin{center}
\begin{turn}{0}
\mbox{\epsfig{file=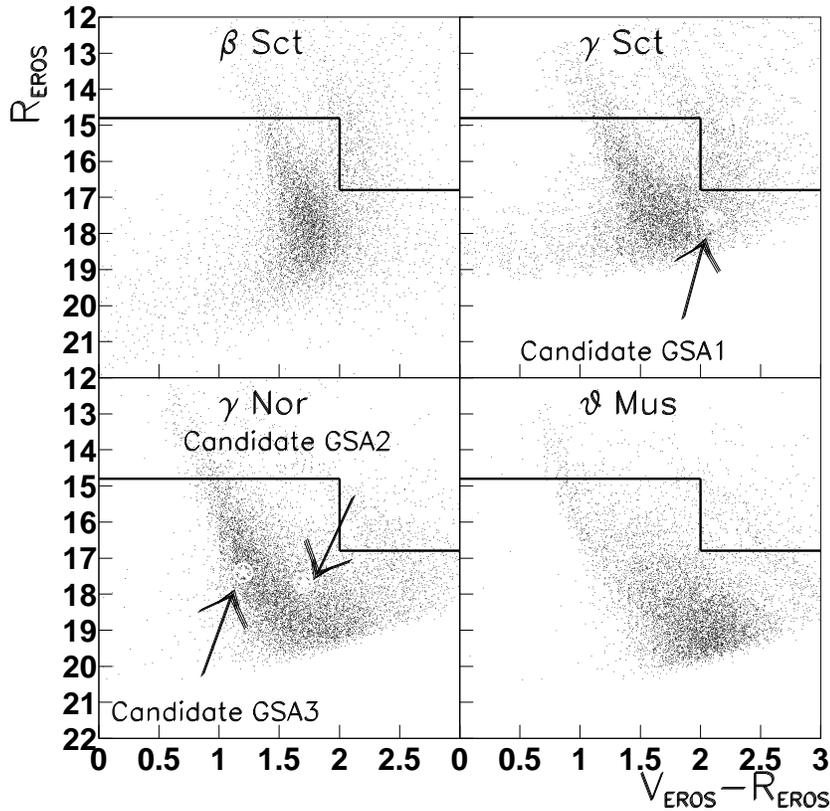,width=12.cm}}
\end{turn}
\caption[]{Colour-magnitude diagrams ($R_{EROS}$ vs $V_{EROS}-R_{EROS}$)
for the stars monitored by {\sc EROS} in
the directions of $\beta$ \& $\gamma$ Sct, $\gamma$ Nor and $\theta$ Mus.
The boxes drawn in the upper corners correspond to the zones excluded
from the search. The positions of the 3 candidates are indicated.}
\label{diagHR}
\end{center}
\end{figure*}
\begin{figure}
\begin{center}
\begin{turn}{0}
\mbox{\epsfig{file=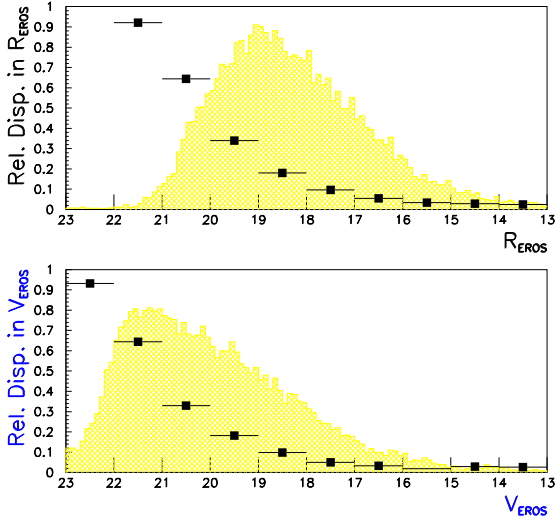,width=8.cm}}
\end{turn}
\caption[]{Relative frame to frame average dispersion of the luminosity
measurements versus $R_{EROS}$ (upper panel) and $V_{EROS}$  
(lower panel), for stars with at least 50 reliable
measurements for each colour.
This dispersion is taken as an estimator of the mean photometric precision.
The superimposed hatched histograms show the magnitude distribution of
the stars in {\sc EROS} bands.}
\label{fg4}
\end{center}
\end{figure}
\begin{figure}
\begin{center}
\begin{turn}{-90}
\mbox{\epsfig{file=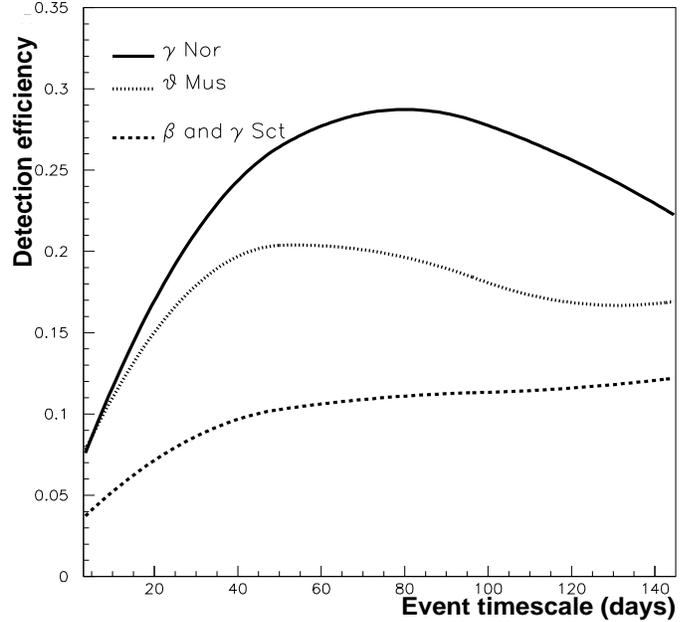,width=9.5cm}}
\end{turn}
\caption[]{{\sc EROS} microlensing detection efficiency 
$\epsilon(\Delta t)$
as a function of $\Delta t$, the Einstein radius crossing time.
$\epsilon(\Delta t)$ is the ratio of
the number of simulated events, satisfying the selection criteria 
with duration $\Delta t$, with
any $u_{0}$ and any date of peak magnification within the research
period (650 days),
to the number of events generated with $u_{0}\le 1$.}
\label{efficiency}
\end{center}
\end{figure}
\section{The search for lensed stars}
\begin{figure}[h!]
\begin{center}
\includegraphics[height=6.cm,width=9.5cm,angle=0,draft=false]{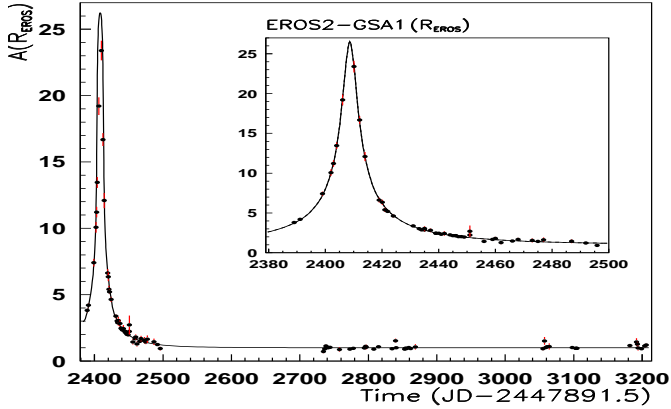}
\includegraphics[height=6.cm,width=9.5cm,angle=0,draft=false]{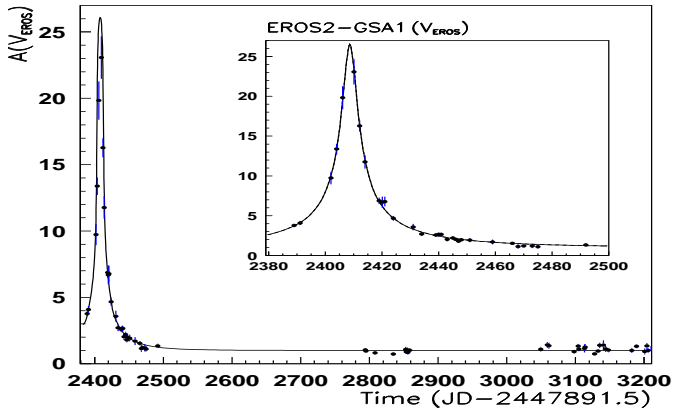}
\caption[]{Magnification curves of the microlensing candidate GSA1
in the direction of $\gamma\; Sct$. The fitted standard microlensing
curve is superimposed (solid line).}
\label{fg6a}
\end{center}
\end{figure}
\begin{figure}[h!]
\begin{center}
\includegraphics[height=6.cm,width=9.5cm,angle=0,draft=false]{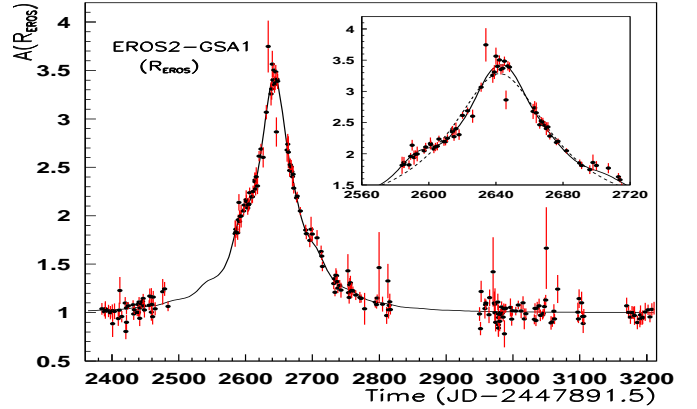}
\includegraphics[height=6.cm,width=9.5cm,angle=0,draft=false]{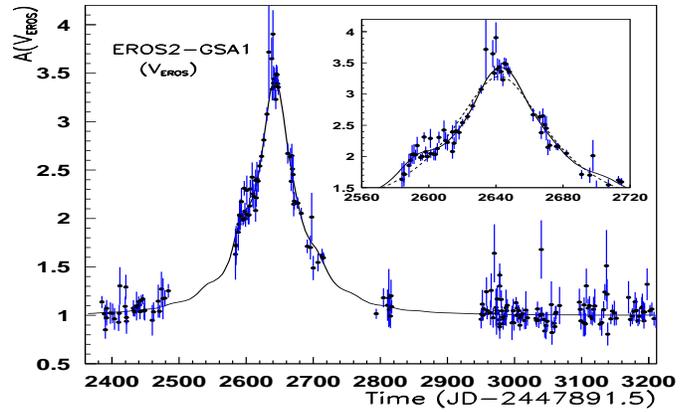}
\caption[]{Magnification curves of the microlensing candidate GSA2
in the direction of $\gamma\; Nor$. The solid line shows
the fitted microlensing curve taking into account the modulation due to a
dominant source orbiting in a binary system with period $P_0=98$ days.
The dashed line corresponds to the best standard microlensing fit.}
\label{fg6b} 
\end{center}
\end{figure}
\begin{figure}[h]
\begin{center}
\includegraphics[height=6.cm,width=9.5cm,angle=0,draft=false]{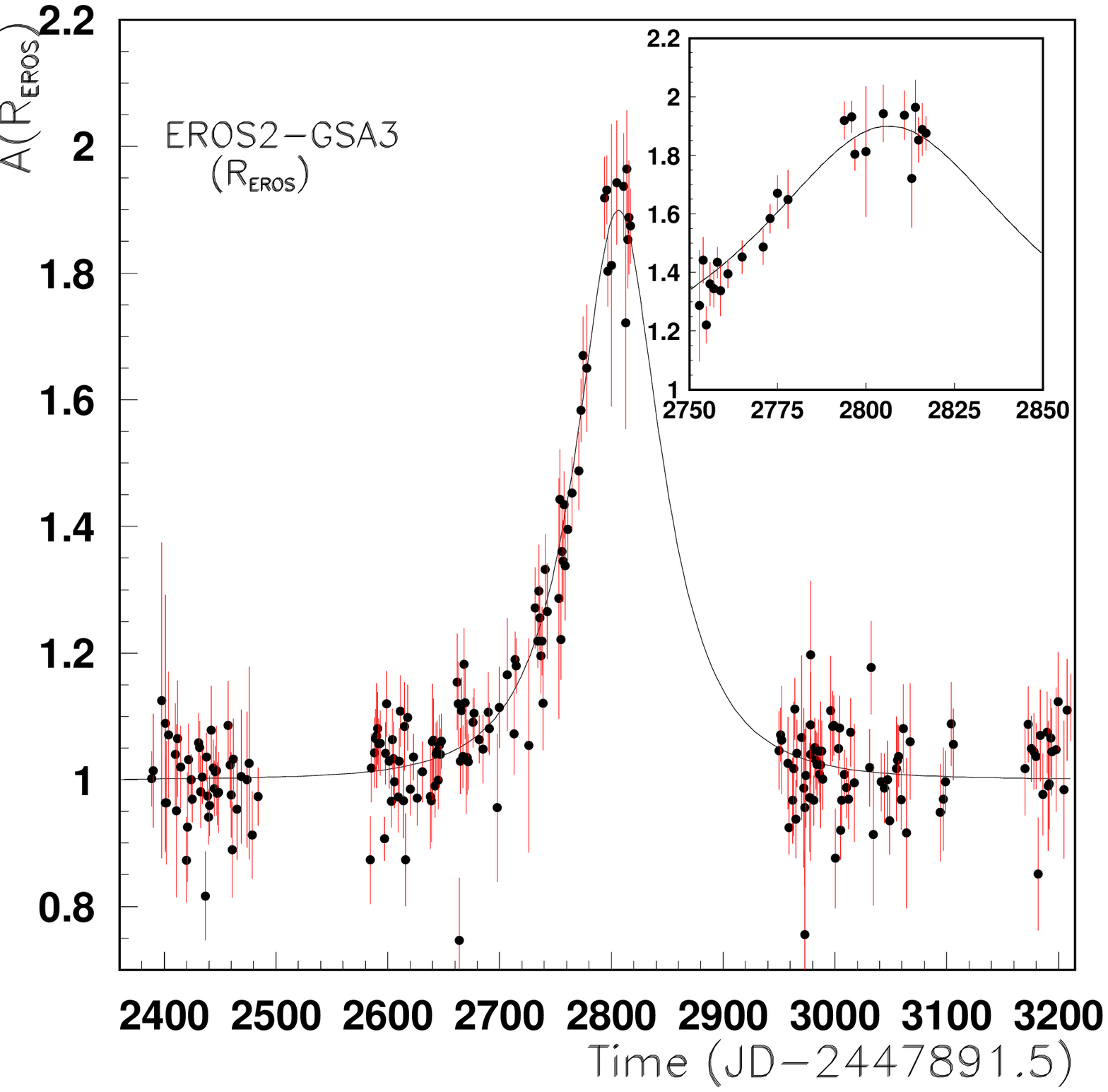}
\includegraphics[height=6.cm,width=9.5cm,angle=0,draft=false]{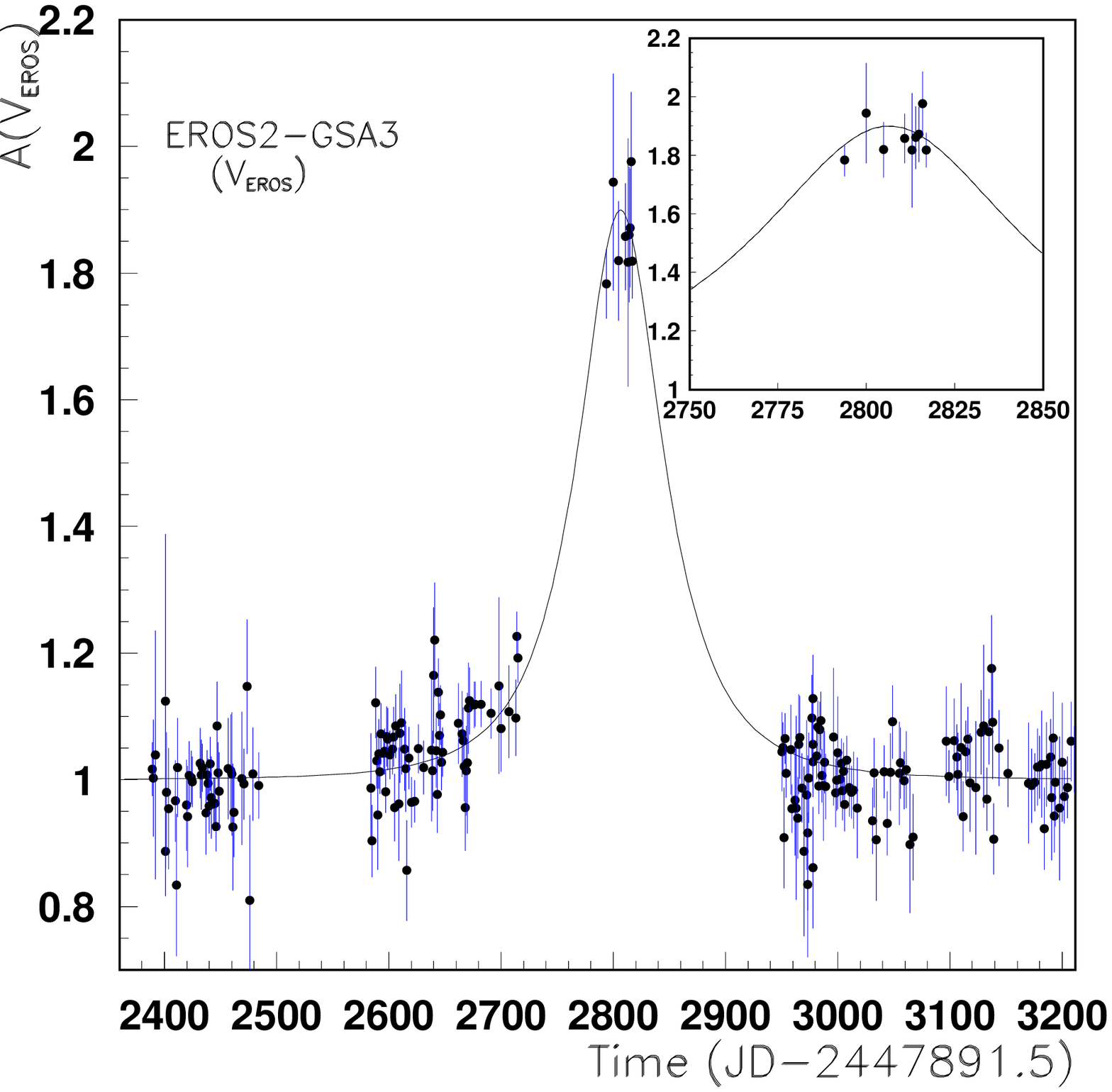}
\caption[]{Magnification curves of the microlensing candidate GSA3,
in the direction of $\gamma\;Nor$.}
\label{fg6c}
\end{center}
\end{figure}
\subsection{Data processing and analysis}
Light curves have been produced from the sequences of images using
the specific software {\sc PEIDA} ({\it Photom\'etrie et
\'Etude d'Images Destin\'ees \`a l'Astrophysique}),
designed to extract photometric information in crowded fields
(\cite{PEIDA}).
Figure \ref{fg4} shows the mean point-to-point relative dispersion
of the measured fluxes along those light curves as a function
of $R_{EROS}$ and $V_{EROS}$.
A systematic search for microlensing events has been performed 
on a set of about 300,000 stars per field 
($8.5 \times 10^{6}$ light curves in total), 
with an average of 70 measurements in the $V_{EROS}$ and $R_{EROS}$
colours (between July 1996 and November 1997).

This search is based on procedures and criteria similar to the ones
developed for the {\sc EROS} microlensing search in the 
{\sc SMC} data (\cite{ErosSMC}).
These criteria are designed to select microlensing events using the
expected characteristics of their light curves
(one single peak, simultaneous in the visible and red bands, with
a known shape), and to reject variable stars
(rejection of stars lying in regions of the colour-magnitude
diagrams mostly populated by variable stars, and
requirement for the stability of the curve outside the peak).
They are also tuned to be loose enough to avoid rejection of non-standard
microlensing events, which have a different peak shape.

The main difference from the {\sc SMC} analysis is the rejection against
variable stars from the instability strip and against red giant variables.
Unlike the Magellanic Clouds case where the positions of these populations
in the colour-magnitude diagram are known {\it a priori},
the scatter and imperfect knowledge of the distances and reddenings of
our target stars make our colour-magnitude cut somewhat empirical.
In particular, its acceptance is different from one field to another
(see the excluded regions in Fig. \ref{diagHR}).
\subsection{The efficiency of the analysis}
To determine the efficiency of each selection criterion, we
have applied them to Monte-Carlo generated light curves,
obtained from a representative sample of the observed
light curves, on which we superimpose randomly generated
microlensing effects.
The microlensing parameters are uniformly drawn in the
following intervals:
impact parameter expressed in Einstein radius unit $u_0\,\in \,[0,2]$,
maximum magnification time in a research period $T_{obs}$
starting 50 days before
the first observations and ending 50 days after the last 
observations\footnote{Outside this
domain of parameters, the detection efficiency is
negligible.}, and Einstein radius crossing time $\Delta t\in [1,150]$ days.
The analysis efficiency (or sampling efficiency) shown in
Fig. \ref{efficiency} is relative to a set of unblended stars.
Blending of a lensed star results in the mixing of unmagnified
light with the expected light curve. The efficiency of the
search and the reconstruction of the lensing parameters
are hence affected by this distortion. On the other hand,
the set of stars that could undergo a detectable lensing effect
is larger than the sample of the monitored stars, and the
expected number of events is also affected.
These effects, which are usually included as corrections to
the sampling efficiency, are not taken into account in
the present article and will require a specific study.
For other targets ({\sc SMC}, {\sc LMC}), we have determined the
correction of the detection rate due to
blending to be less than 10-20\% (\cite{TheseRenault}; \cite{ThesePalanque}).
\subsection{Results of the selection}
Three light curves satisfy all the requirements and are
hereafter named candidates and labelled GSA1 to 3.
Figures \ref{fg6a}, \ref{fg6b} and \ref{fg6c} show the light curves of each
candidate and Table \ref{tab2} contains their characteristics.
Measurements taken after Jan 1st, 1998 (date 2922) are shown, although they
were not used in the selection.
\begin{table*}
\caption[]{Characteristics of the 3 microlensing candidates}
\begin{center}
\begin{tabular}{llll}
\hline\noalign{\smallskip}
Candidate & EROS2-GSA1	& EROS2-GSA2	& EROS2-GSA3 \\
\noalign{\smallskip}
\hline\noalign{\smallskip}
field & $\gamma\; Sct$  & $\gamma\; Nor$ & $\gamma\; Nor$ \\
Coordinates of star (J2000)	& $\alpha=18{\rm h}29{\rm m}09.0{\rm s}$	& $\alpha=16{\rm h}11{\rm m}50.2{\rm s}$	& $\alpha=16{\rm h}16{\rm m}26.7{\rm s}$	\\
				& $\delta=-14\degree15'09"$& $\delta=-52\degree56'49"$& $\delta=-54\degree37'49"$\\
Galactic coordinates		& $b=-1.\hskip-2pt \degree 67$	& $b=-1.\hskip-2pt \degree 17$	& $b=-2.\hskip-2pt \degree 86$	\\
				& $l=17.\hskip-2pt \degree 75$	& $l=330.\hskip-2pt \degree 49$	& $l=329.\hskip-2pt \degree 82$	\\
$R_{EROS}$	& $17.6$		& $17.7$		& $17.4$\\
$V_{EROS}$	& $19.7$		& $19.4$		& $18.6$\\
Date of maximum magnification	& Aug. $3^{rd}$, 1996		& Mar. $26^{th}$, 1997		& Oct. $7^{th}$, 1997\\
Julian Day -21,447,891.5	& $2408.6 \pm 0.1$		& $2642.9 \pm 0.2$		& $2806.2 \pm 1.1$\\
Einstein radius crossing & $73.5\pm 1.4$	& $98.3\pm 0.9$	& $70.0\pm 2.0$	\\
time $\Delta t = R_E/V_T$ (days)	\\
Max. magnification	& $26.5\pm 0.6$	& $3.05\pm 0.02$	& $1.89\pm 0.01$	\\
Impact parameter (in $R_E$)	& $0.0378\pm 0.001$	& $0.342\pm 0.002$	& $0.593\pm 0.007$	\\
$\chi^2$ of best fit	& 185.7/163 d.o.f.	& 551/425 d.o.f.	& 445/427 d.o.f.	\\
							\\
Remarks		& $M_{lens} > 4.6 \times 10^{-3}\Msol$	& binary source fit	& \\
		& at 95\% C.L. (see text)			& period 98 days (see text)	& \\
\noalign{\smallskip}
\hline
\end{tabular}
\end{center}
\label{tab2}
\end{table*}
\section{Optical depth and event timescales}
\subsection{Optical depth}
The optical depth towards a pointlike source is defined 
as the fraction of time during which it undergoes a lensing
magnification larger than 1.34.
For a given target the measured optical depth $\tau$
is computed from:
\begin{eqnarray*}
\tau &=& \frac{1}{N_{obs}T_{obs}}\frac{\pi}{2}\sum_{events}
\frac{\Delta t}{\epsilon (\Delta t )} \ , 
\label{tau}
\end{eqnarray*}
where $N_{obs}$ is the number of monitored stars in
the target, $T_{obs}$ is the duration of the search period
(650 days for this 2 year analysis)
and $\epsilon (\Delta t)$ is the average detection efficiency
normalized to the microlensing events with impact parameter $u_0 < 1$,
whose maximum magnification takes place within the research period.
The contribution of the candidates to the optical depth
is given in Table 3.
\begin{table}
\begin{center}
\caption[]{Contribution of the candidates to the optical depth $\tau$
assuming the sources to be $7\ \rm{kpc}$ away. 
In the case of $\theta$ Mus we give a 95\% C.L. upper limit on the optical
depth contribution from events with $\Delta t = 80$ days
($\epsilon(80\ {\rm days})$ = 18\%). }
\begin{tabular}{|l|cc|c|c|}
\hline
Target & \multicolumn{2}{|c|}{Scutum}  & Norma & Musca \\
Direction & $ \beta\; Sct$  & $\gamma\; Sct$ & $\gamma\; Nor$ & $\theta\; Mus$\\
\hline
$<b>$		& $-2.\hskip-2pt \degree 5$ & $-2.\hskip-2pt \degree 6$ & $-2.\hskip-2pt \degree 7$ & $-1.\hskip-2pt \degree 8$ \\
$<l>$		& $27.\hskip-2pt\degree 0$ & $18.\hskip-2pt \degree 6$ & $331.\hskip-2pt \degree 2$ & $306.\hskip-2pt \degree 4$ \\
Events detected	& none	& GSA1	& GSA2 \& 3 & none	\\
$\Delta t$ (days)& -	& $73$	& $98$ \& $70$ & (80) \\
$\epsilon(\Delta t)$(\%) 	& \multicolumn{2}{|c|}{$10$}	& $27$	& (18) \\
$\tau$(event)($\times 10^{6}$)	& - & 1.02 & $0.29\ \&\ 0.21$ & -\\

$\bar\tau$(target)($\times 10^{6}$) &\multicolumn{2}{|c|}{$0.47$} & $0.5$ &$<1.82$ at \\
		   &\multicolumn{2}{|c|}{      } & 	 &95\% CL \\
\hline 
\multicolumn{4}{|c|}{$\bar\tau $ {\rm averaged \ over \ the \ 4 \ directions} ($\times 10^{6}$)} & $0.38^{+0.53}_{-0.15}$\\

\hline
\end{tabular} 
\end{center}
\label{tab3} 
\end{table}

We have modeled the Galaxy in two different ways using three components:
a central bulge, a disc and a dark halo.
The density distribution for the bulge -~a barlike triaxial model~-
is taken from \cite{Dwek} model G2 (in cartesian coordinates):
\begin{eqnarray*}
\rho_{Bulge} &=&  \frac{M_{Bulge}}{8 \pi abc} e^{-r^{2}/2} \ , \ 
r^{4} = \left[ \left( \frac{x}{a} \right)^{2} + 
	       \left( \frac{y}{b} \right)^{2} \right]^{2} + 
	\frac{z^{4}}{c^{4}} \ ,
\end{eqnarray*}
where $M_{Bulge}$ is the bulge mass, and a, b, c are the length
scale factors.
The bar major axis is inclined at an angle of $15\degree$ 
with respect to the Sun-Galactic Centre line.

We compare our measurements with the predictions of two models
with extreme disk contributions (here the main structure involved
in microlensing).
The first one has a ``thin'' disc and a standard isotropic and
isothermal halo (Model 1) with a density distribution given in
spherical coordinates by:
\begin{eqnarray*}
\rho(r) = \rho_{h\odot} \frac{R_{\odot}^{2}+R_{c}^{2}}{r^{2}+R_{c}^{2}}\ ,
\end{eqnarray*}
where $\rho_{h\odot}$ is the local halo density,
$R_{\odot}=8.5\ \rm{kpc}$ is the distance between the Sun and the
Galactic Centre, and
$R_{c} = 5\ \rm{kpc}$ is the Halo ``core radius''.
The matter distribution in the disc is modeled in cylindrical
coordinates by a double exponential (see e.g
\cite{Bienayme} and \cite{Schaeffer}):
\begin{eqnarray*}
\rho_{thin}(R,z) = \frac{\Sigma_{thin}}{2H_{thin}} \exp 
\left(\frac{-(R-R_{\odot})}{R_{thin}} \right) \exp 
\left( \frac{-|z|}{H_{thin}} \right) \ ,
\end{eqnarray*}
$\Sigma_{thin}$ is the column density of the disc at the Sun position,
$H_{thin}$ is the height scale 
and $R_{thin}$ is the length scale of the disc.

The second model (Model 2) has a ``thin'' and a ``thick'' disc, 
and a very light halo. 
Both models share the same bulge contribution.
The model parameters are summarized in Table 4.
\begin{table}
\begin{center}
\caption[]{Parameters of the galactic models used in this article.}
\begin{tabular}{|c|l|c|c|}
\hline
	& Parameter			&Model 1&Model 2 \\ \hline
	& $a\ ({\rm kpc)}$	& \multicolumn{2}{c|}{1.49}	\\ 
Bulge	& $b\ ({\rm kpc})$	& \multicolumn{2}{c|}{0.58}	\\ 
	& $c\ ({\rm kpc})$    & \multicolumn{2}{c|}{0.40}	\\
	& $M_{bulge} (\times 10^{10}M_{\odot})$	& \multicolumn{2}{c|}{2.1} \\ \hline
	& $\Sigma\ (M_{\odot} {\rm pc}^{-2})$	& \multicolumn{2}{c|}{50} \\ 
Thin disc	& $H\ ({\rm kpc})$   	              	& \multicolumn{2}{c|}{0.325} \\ 
	& $R\ ({\rm kpc})$		        & \multicolumn{2}{c|}{3.5}  \\
	& $M_{thin}(\times 10^{10}M_{\odot})$	& \multicolumn{2}{c|}{4.3} \\ \hline
	& $\Sigma\ (M_{\odot} {\rm pc}^{-2})$ & -    & 35      \\ 
Thick disc	& $H\ ({\rm kpc})$	                & -	& 1.0     \\ 
	& $R\ ({\rm kpc})$	                & -    & 3.5   \\
	& $M_{thick}(\times 10^{10}M_{\odot})$	& -	& 3.1 \\ \hline
	& $\rho_{h\odot}\ (M_{\odot} {\rm pc}^{-3})$ & 0.008	& 0.003  \\
Halo	& $R_{c}\ ({\rm kpc})$                & 5.0  & 5.0    \\ 
	& $M\ in\ 60 \ {\rm kpc}\ (10^{10} M_{\odot})$   & 51	& 7     \\ \hline
\hline
	& $\rho_{\odot}\ (M_{\odot} {\rm pc}^{-3})$	& 0.085	& 0.098	\\
Predictions	& $V_{rot}\ at\ sun\ ({\rm km}\ {\rm s}^{-1})$   & 211  & 222    \\ 
	& $V_{rot}\ at\ 20\ {\rm kpc}$   & 203  & 180     \\
	& $V_{rot}\ at\ 60\ {\rm kpc}$	 & 200  & 140     \\ \hline       
\noalign{\smallskip}
\end{tabular}
\end{center}
\label{tabmodel}
\end{table}
Fig. \ref{optvsl} shows the expected optical depth up to $7\ {\rm kpc}$
as a function of longitude for both models, at the average
latitude of our fields $b=-2.5\degree$.
As the main contribution comes from the thin disc (about 90\%),
variations of the optical depth from field to field
due to the range of $2$ to $3\degree$ in latitude can reach $\simeq 30\%$
in the case of $\gamma$ Nor, and $\simeq 20\%$ for the other targets.

The expected optical depth, averaged over the four directions, is 
$0.60 \times 10^{-6}$ for model 1, and $0.70 \times 10^{-6}$ for model 2.
These estimates vary by 50\% if the average distance of the sources
is changed by 2kpc, or if the parameter $\Sigma_{thin}$ is changed by
$25\Msol/{\rm pc^2}$.
The measured optical depth, averaged over the four directions is: 
\begin{eqnarray*}
\bar\tau = 0.38^{+0.53}_{-0.15} \times 10^{-6}.
\end{eqnarray*}
The confidence interval reported here takes into account Poisson
fluctuations and the possible event timescale variations
inside the range $[71,98]$ days.
The comparison of Fig. \ref{optvsl} with Table 3 also shows that
the measured optical depths in the four directions are compatible with
the predictions of both models. 
\subsection{Microlensing event timescales}
The duration of the three events is long
($\sim 80$ days in average). 
Fig. \ref{duration} shows the expected event duration 
distribution towards $\gamma\ Nor$ within the framework of model 1.
This distribution is obtained assuming the following mass functions
and kinematical characteristics:
\begin{itemize}
\item
We assume that lenses belonging to the (non-rotating) halo
have the same mass ($0.5 M_{\odot}$); their
velocities transverse to the line of sight of disc stars follow
a Boltzmann distribution with a dispersion of $\sim$~$150{\rm km}/{\rm s}$.
Thus the expected duration of microlensing events is small
(see Fig. \ref{duration}).
\item
For lenses belonging to the bulge, the mass function is taken from
\cite{richer}
and the velocities transverse to the line of sight of disc stars
also follow
a Boltzmann distribution with a dispersion of $\sim$~$110{\rm km}/{\rm s}$.
The expected rate of microlensing events due to this structure towards
the direction considered in Fig. \ref{duration} ($\gamma Nor$)
is found to be negligible.
\item
The disc lenses mass function is taken from \cite{Gould-1997}, which
is derived from {\sc HST} observations.
Disc lenses are subject to a similar global rotation as the observer
and the sources (\cite{rotdisc}).
We assume a negligible particular motion of the monitored sources
with respect to the spiral arms, because they are probably young stars. 
Following \cite{Griest}, the motion of
the Sun relative to the Local Standard of Rest is taken as
$(v_{\odot R}=9,\ v_{\odot \theta}=11,\ v_{\odot z}=16)$ (in {\rm km}/{\rm s})~;
the velocity dispersions of the lens population are expected to be
$(\sigma(V_R)=40,\ \sigma(V_{\theta})=30,\ \sigma(V_z)=20)$ (in {\rm km}/{\rm s}).
As the disc lenses have a low velocity relative to the line of sight,
disc-disc events have longer timescales, as can be seen on Fig. \ref{duration}.
\end{itemize}

Assuming that all three lenses
belong to the halo leads to large probable masses
($> 2 \Msol$). 
As such high mass lenses would be visible stars that we
do not observe (dismissing the unlikely possibility that the three
could be neutron stars), the disc-disc lensing hypothesis is more probable.
\begin{figure}
\begin{center}
\begin{turn}{0}
\mbox{\epsfig{file=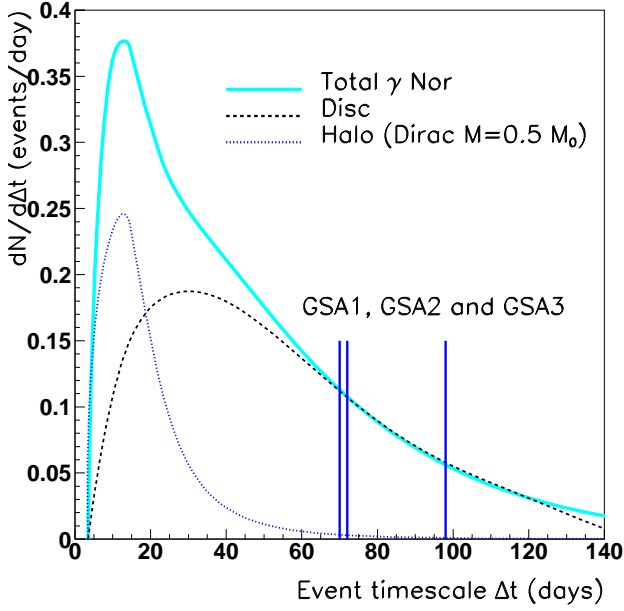,width=8.8cm}}
\end{turn}
\caption[]{$\Delta t$ distribution
of the events expected during the search
period $T_{obs}$ per $10^7$ monitored stars towards
$\gamma\ Nor$, within model 1.
This distribution takes into account
the detection efficiency $\epsilon(\Delta t)$.
The durations of the three candidates are marked.
Notice that the bulge contribution is negligible.
\label{duration}}
\end{center}
\end{figure}
This hypothesis could also explain
the fact that 4 events over the 45 found by the {\sc MACHO}
Collaboration towards the Galactic Centre (see \cite{MachoBulge-1an})
have a timescale $>50$ days, significantly larger
than the mean for the whole sample (21 days).
Indeed, the contribution of these 4 events to the total optical depth
is about $0.6 \times 10^{-6}$, which is compatible with the expectation
of the disc lensing contribution ($\sim 0.7\times 10^{-6}$ for Model 1).
\section{Detailed analyses of the candidates}
Detailed analyses have been performed on
GSA1 and GSA2 candidates, whose light curves are sufficiently
sampled. For this purpose, we have used all available
data on the candidates (i.e. three years of observations).
\subsection{Candidate EROS2-GSA1}
Candidate GSA1
exhibits a large magnification ($A_{peak}>25$ at 95\% C.L.)
with no detectable blending\footnote{Fits including more than 10\%
blending are excluded at 95\% C.L. Moreover, since
the magnification curves are the same in both colours,
the amplified star and an hypothetical blending star
would have to have -by chance- the same colour.} 
and no detectable parallax effect (\cite{GouldParallax}).
Despite the small impact
parameter, no evidence for a distortion of the curve
due to the non-zero size of the lensed source is found. 
A fitting procedure allows one to put an upper limit to the ratio
of the angular stellar radius $\theta_{\star}$ to the angular
Einstein radius $\theta_E$ of the lens:
\begin{eqnarray*}
\frac{\theta_{*}}{\theta_E} = \frac{\theta_{*}}{\sqrt{\frac{4GM}{c^2}\frac{1-x}{xD}}} < 0.066  \ at \ 95\% \ C.L\ ,
\end{eqnarray*}
where $M$ is the lens mass, 
$D$ the distance of the observer to the source and $xD$ its
distance to the lens.
From the position of the source in the colour-magnitude diagram we can
assume that its temperature is comparable to or
lower than that of the Sun. We then obtain:
\begin{eqnarray*}
\frac{\theta_{\star}}{\theta_{\odot}\ {\rm at}\ 10\ {\rm pc}}\ >
10^{(V_{EROS\odot}^{abs.}-V_{EROS\star}^{app.})/5}\ ,
\end{eqnarray*}
where the apparent magnitude $V_{EROS\star}^{app.}$ should be corrected
for the interstellar extinction.
Ignoring this correction leads to the conservative
limit $\theta_{\star} > 2.37\times 10^{-12} \ {\rm rad}$,
implying that $\theta_E > 3.6 \times 10^{-11} \ {\rm rad}$
and that the angular proper motion of the deflector:
\begin{eqnarray*}
\mu\ ({\rm km/s/kpc})\ = 3.57\times10^{11}\frac{\theta_E({\rm rad})}{\Delta t\ ({\rm days})} > 0.17\ at \ 95\% \ C.L.
\end{eqnarray*}
Assuming the interstellar extinction
to be 3 magnitudes (see Sect. 2.2) leads 
to $\theta_{\star} > 9.42\times 10^{-12} \ {\rm rad}$,
implying that $\theta_E > 1.43\times 10^{-10} \ {\rm rad}$
and: 
\begin{eqnarray*}
\mu > 0.69\ {\rm km/s/kpc} \ at \ 95\% \ C.L.
\end{eqnarray*}
Since the typical expected value (based on Galactic kinematics)
is $\mu \simeq 4\ {\rm km/s/kpc}$, this limit does not probe the expected
range of parameter space.

On the other hand, the standard microlensing fit is satisfactory, and
taking into account the possibility of a parallax effect does
not result in a significantly better fit.
We can then infer a lower limit on $\tilde r_E$, the size of the
Einstein ring projected from the source onto the observer's plane:
\begin{eqnarray*}
\tilde r_E = \frac{\theta_E xD}{(1-x)} > 1.33\ {\rm AU}\ at\ 95\% CL.
\end{eqnarray*}
This implies a lower limit on the projected speed of the lens onto the
local transverse plane: 
\begin{eqnarray*}
\tilde v = \tilde r_E/\Delta t > 31.6\ {\rm km/s} \ at \ 95\% \ C.L.
\end{eqnarray*}
Since typical values expected for this quantity are
of the order of $\tilde v \simeq 100\ {\rm km/s}$,
it is clear that this is not a very constraining limit either.
Fig. \ref{parall} shows the excluded area for the tip of the
2-dim. vector $\overrightarrow{\tilde{v}}$
in the local transverse plane. The lower limit we find for
the modulus $\tilde v$
(and $\tilde r_E$) is relatively small, because
the best parallax fit, which is not
significantly better than a standard fit (183.7/161 d.o.f. compared
with 185.7/163 d.o.f.), is obtained for
$\overrightarrow{\tilde v} = (-17.1,-31.1)\ {\rm km/s}$
in the frame of Fig. \ref{parall}.
This special configuration produces a
distorted light curve which
diverges only marginally from the standard fitted curve, and
only during periods where the measurements are not very precise.

From the definitions of $\theta_E$ and $\tilde r_E$ one gets
the relation:
\begin{eqnarray*}
M = \frac{c^2}{4G}\tilde{r}_E\theta_E = \frac{c^2}{4G}\tilde v\mu\Delta t^2\ ,
\end{eqnarray*}
from which we derive a lower limit on the lens mass by combining the
two constraints from finite size
and parallax analysis, i.e. at 95\% C.L.:
\begin{eqnarray*}
M & > & 1.2 \times 10^{-3} \Msol 
\ \mathrm{\ ignoring \ interstellar \ extinction}, \nonumber \\
M & > & 4.6 \times 10^{-3} \Msol 
\ \mathrm{\ with\ 3\ magnitudes\ of\ extinction}. \nonumber
\end{eqnarray*}
Note that this limit is independent of the lens and source distances
to the observer.
The fact that this lower limit is so much smaller than the expected
mass for such lenses follows immediately from the fact that each
limit from which it is derived is not probing the expected range
of kinematic parameters.
Fig. \ref{mdvsx} shows the excluded areas in the $M$ versus $x$ plane,
from the finite size study 
and the parallax analysis, 
assuming the source to be located $7$ {\rm kpc} away.
Exclusion curves for the
two hypotheses on the interstellar absorption are shown. 
A better characterization of the lensed source should allow one to
refine these preliminary studies.

Finally, a limit on the lens luminosity can be derived from the
maximum blending limit allowed by the fit. At 95\% C.L.,
the apparent magnitude of the lens is at least 2.5 magnitudes above the 
measured baseline magnitude, i.e. $R_{EROS}>20.1$ and
$V_{EROS}>22.2$. 

\begin{figure}
\begin{center}
\mbox{\epsfig{file=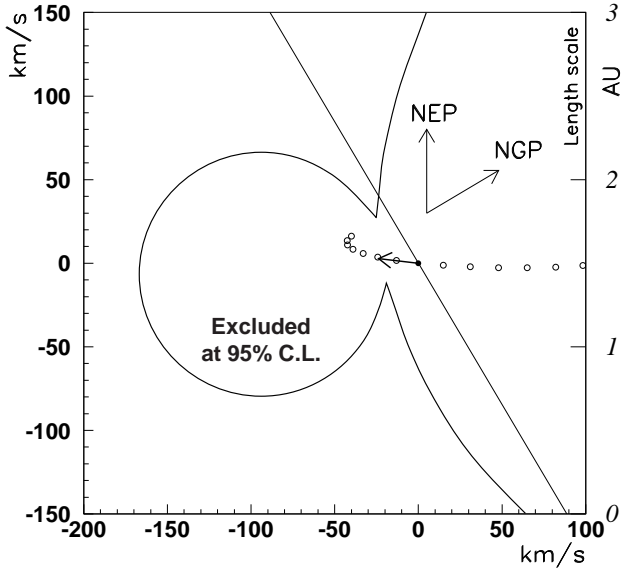,width=9.5cm,angle=0}}
\caption[]{Excluded area for the tip of the 2-dim. vector
$\overrightarrow{\tilde v}$ in the local transverse plane.
The smallest speed compatible with the observations
corresponds to a deflector with a velocity
oriented towards the lower spike.
The arrow shows the earth projected velocity at the maximum magnification
time.
The straight line is the intersection of the Galactic plane.
The projection of the earth trajectory is indicated by the open circles,
starting 60 days before maximum, ending 70 days after maximum,
with 10 days spacing (right scale in AU).
{\sc NEP} is the North Ecliptic Pole direction, {\sc NGP} is the North
Galactic Pole direction.
\\
\label{parall}}
\end{center}
\end{figure}
\begin{figure}
\begin{center}
\mbox{\epsfig{file=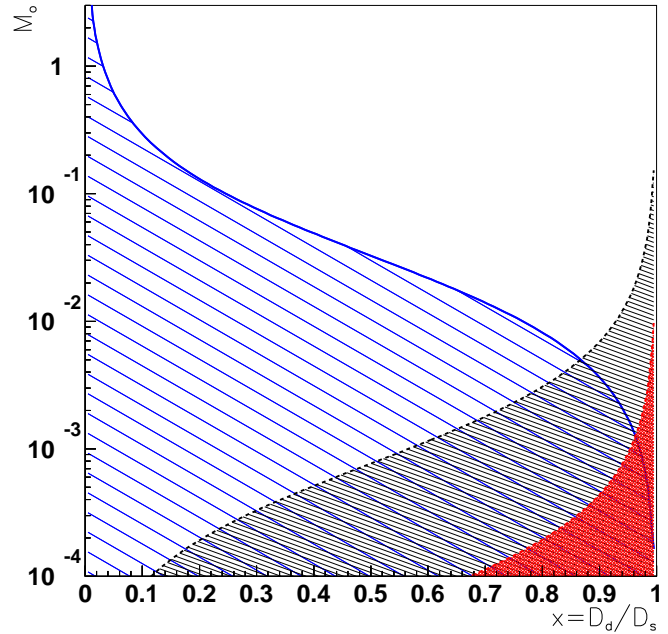,width=9.5cm}}
\caption[]{95\% C.L. exclusion diagram from the parallax analysis
and from the finite source size study for the candidate EROS2-GSA1.
The lower curve in the right part is the conservative limit
ignoring interstellar absorption, the higher curve corresponds
to a 3 magnitude absorption.
These curves assume the source to be located $7\ {\rm kpc}$
away. If the source is assumed to be at some other distance $D$, one
has to scale the increasing functions by the factor $D/(7\ {\rm kpc})$,
and the decreasing function by $(7\ {\rm kpc})/D$.
\\
\label{mdvsx}}
\end{center}
\end{figure}
\subsection{Candidate EROS2-GSA2}
GSA2 has residuals to a standard microlensing fit which clearly
exhibit a modulation; a period of $\sim 54$ days is found in the residuals 
of the standard fit during the magnification.
On the other hand, taking into account a possible parallax effect does not 
significantly improve this fit.
Furthermore, we know that the lensing of a periodic variable
star would produce a modulation of the light curve whose amplitude
should follow the magnification. 
As we do not detect a significant
modulation in the non-magnified part of the light curve,
the most probable origin of this modulation is that one
dominant luminous source or two sources orbit around
the centre of gravity of a binary system, inducing a wobbling of
the line of sight with respect to the trajectory of the lens
\footnote{This type of configuration has been studied by
\cite{SourceBin}, \cite{sazhin} and \cite{HanBin},
and was already mentioned in our earlier article (\cite{variab}).}.
In this configuration, a modulation of the light curve is expected
to occur only during the lensing peak. We obtain equally satisfactory
fits for two classes of circular binary systems
(with at best $\chi^2=551$ for 425 d.o.f.,
to be compared with 876 for 439 d.o.f. for a standard microlensing fit).

The first class includes systems with a dominant source orbiting with
a period $P_o \sim 51$ days,
and with the projected semi-axis $\rho=ax/R_E \sim 0.04$
\footnote{Fits which are slightly less good can be obtained
with larger periods,
but in this case only the first shoulder (around date=2600) is correctly
matched.}.
The second class corresponds to systems with two luminous stars
orbiting with a period 
$P_o \sim 98$ days, with luminosity and mass ratios around
1/3-1/2, and a projected distance between the two components of about
$\rho=ax/R_E \simeq 0.4$.
Given the domain of values for $x$ and $R_E$, these parameters
are compatible with physically acceptable masses of the
binary satisfying Kepler's third law.
Many more configurations can fit the observations
if we consider systems with elliptic orbits.
\begin{figure}
\begin{center} 
\mbox{\epsfig{file=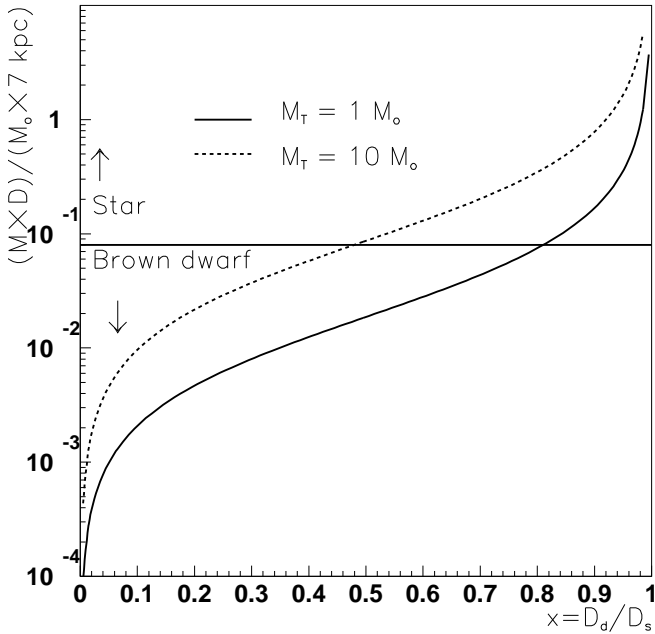,width=9.5cm}}
\caption[]{Relation between $M_{lens}\times D$ and $x$ assuming 
a binary system with $P_o = 100\ days$, $\rho = 0.4$ and a total mass
$M_T$ of $1 \Msol$ or $10 \Msol$.
\\
\label{gnexclu}}
\end{center}
\end{figure}
A spectroscopic study of the source is under way in order to test
its binarity. If the source proves to be a spectroscopic binary,
then one can get a good estimate or constraint on the angular Einstein radius
$\theta_E$ (and then on the angular proper motion of the lens $\mu$).

Expressing the Einstein radius and Kepler's third
law in the relation $\rho=ax/R_E$ leads to:
\begin{eqnarray*}
\frac{M_{lens}}{M_{\odot}}\frac{D}{7 \ {\rm kpc}} \sim
	6.7.10^{-6} \left[ \frac{P_o}{1 \ {\rm day}} \right]^{\frac{4}{3}}
	\left[ \frac{M_{T}}{M_{\odot}} \right]^{\frac{2}{3}} \rho^{-2}
 	\frac{x}{1-x}\ .
\end{eqnarray*}
Fig. \ref{gnexclu} illustrates this relation between $M_{lens}\times D$ and $x$
assuming the source to be a binary system with $P_o = 100$ days, with 
$\rho = 0.4$ and with two different hypotheses for the
mass $M_T$ of the system.

\section{Conclusion}
We have searched for microlensing events with durations
ranging from a few days to a few months in four Galactic disc
fields lying $18^\circ$ to $55^\circ$ from the Galactic Centre.
We find three events
that can be interpreted as microlensing effects due to
massive compact objects.
Their long duration favours the interpretation of lensing by
objects belonging to the disc instead of the halo.
The average optical depth measured towards the four directions
is $\bar\tau = 0.38^{+0.53}_{-0.15} \times 10^{-6}$.
Assuming the sources to be $7$~${\rm kpc}$ away, the expected optical depths
from two different galactic models vary from $0.60$ to $0.70\times 10^{-6}$,
in agreement with our measurement.

One event displays a modulation of the magnification which is compatible 
with the lensing of a binary source.
More information about the configuration of this possible binary
source, and about the lens mass acting on the strongly amplified
source, are expected from complementary observations.
No evidence for parallax and blending effects has been found.

The observations continues towards the spiral arms.
More accurate measurements should be obtained with the increase of
statistics (\cite{derue}), allowing one to estimate the disc contribution 
to the optical depth towards the bulge and the Magellanic Clouds.
\begin{acknowledgements}
We are grateful to D. Lacroix and the technical staff at the Observatoire
de Haute Provence and to A. Baranne for their help in refurbishing the {\sc MARLY}
telescope and remounting it in La Silla. We are also grateful for
the support given to our project by the technical staff at ESO, La Silla.
We thank J.F. Lecointe for assistance with the online computing.
We wish to thank also C. Nitschelm for his contribution to the data taking.
\end{acknowledgements}

\end{document}